\documentclass[12pt]{article}
\usepackage{amssymb}
\begin{document}

\makeatletter
\@addtoreset{equation}{section}
\renewcommand{\theequation}{\thesection.\arabic{equation}}
\rightline{LAB.UFR-HEP/0009}
\vspace{1.5cm}
\begin{center}
{\bf{\huge Explicit Derivation of \\
Yang-Mills Self-Dual Solutions on \\
non-Commutative Harmonic Space }}
\end{center}
\smallskip
\begin{center}
A. Belhaj$^1$, M. Hssaini$^1$, E.M. Sahraoui$^{1,2}$, E.H. Saidi$^1$\\
\end{center}
\begin{center}
{$^{1}$\it  High Energy Physics Laboratory,  Faculty of Sciences, Rabat, Morocco.}\\
\smallskip
$^2$ {\it Sektion Physik, Universit\"at M\"unchen, Theresienstra$\beta$e 37, 80333 M\"unchen, Germany. }\\
\bigskip
{\tt sahraoui@theorie.physik.uni-muenchen.de}\\
{\tt H-saidi@fsr.ac.ma}
\end{center}

\vspace{1cm}
\def\be{\begin{equation}}
\def\ee{\end{equation}}
\def\bea{\begin{eqnarray}}
\def\eea{\end{eqnarray}}
\def\nn{\nonumber}
\def\hV{\widehat V^{++}}
\def\V{V^{++}}
\def\v{V^{--}}
\def\hv{\widehat V^{--}}
\def\U{U^{++}}
\def\hU{\widehat U^{++}}
\def\u{U^{--}}
\def\hu{\widehat U^{--}}
\def\W{W^{++}}
\def\hW{\widehat W^{++}}
\def\w{W^{--}}
\def\hw{\widehat W^{--}}
\def\D{D^{++}}
\def\hD{\widehat D^{++}}
\def\d{D^{--}}
\def\hd{\widehat D^{--}}
\def\K{K^{++}}
\def\hK{\widehat K^{++}}
\def\k{K^{--}}
\def\hk{\widehat K^{--}}
\def\E{E^{++}}
\def\hE{\widehat E^{++}}
\def\e{E^{--}}
\def\he{\widehat E^{--}}
\def\X{x^{+}}
\def\x{x^{-}}
\def\Y{y^{+}}
\def\y{y^{-}}
\def\f{f^{(q_1)}}
\def\g{g^{(q_2)}}
\def\*{\ast}
\def\[{\bigl[}
\def\]{\bigr]}
\def\({\bigl(}
\def\){\bigr)}
\def\p{\partial}
\def\b{\beta}
\def\a{\alpha}
\def\o{\over}
\def\t{\tau}
\def\ht{\widehat\tau}
\def\ha{\widehat A}
\def\hf{\widehat F}
\def\cd{\cal D}
\def\me{\eta ^{--}}
\def\pe{\eta ^{++}}
\def\S{\Sigma}
\def\cS{\cal S}
\def\R{\bf R}
\def\G{\gamma}
\def\C{C^{++}}
\def\c{C^{--}}
\def\K{K^{++}}
\def\k{K^{--}}
\def\H{H^{++}}
\def\h{H^{--}}
\def\wt{\widetilde}
\begin{center}
{\bf Abstract}
\end{center}

We develop the noncommutative harmonic space (NHS) analysis to study the problem of solving the non-linear constraints eqs of noncommutative  Yang-Mills self-duality in four-dimensions. We show that this space, denoted also as  NHS($\eta,\theta$), has two SU(2) isovector deformations $\eta^{(ij)}$ and $\theta^{(ij)}$ parametrising respectively two noncommutative harmonic subspaces NHS($\eta$,0) and NHS(0,$\theta$) used to study the self-dual and anti self-dual noncommutative Yang-Mills solutions. We reformulate the Yang-Mills self-dual constraint eqs on NHS($\eta$,0) by extending the idea of harmonic analyticity to linearize them. Then we give a  perturbative self dual solution  recovering the ordinary one. Finally we present the explicit computation of an exact self-dual solution.\\

\vspace{2cm}
\noindent{\sf July, 2000}
\newpage
\section{Introduction}

\hspace{1cm}Yang-Mills instantons on noncommutative geometry have been 
subject to some interest nowadays in connection with the recent developments 
in superstrings theory \cite{a} and compactification of matrix model of M-theory 
\cite{b,c,d}. Noncommutative instantons are involved in the study of D(p-4)/Dp
 brane systems (p$\ge 3$) of superstrings; in particular in the ADHM 
construction of the D0/D4  system \cite{e,f,g} and in the 
determination of the
 vacuum solutions of the Higgs branch of supersymmetric gauge theories with 
eight supercharges \cite{h,i,j}.

In string theory, noncommutative geometry appears from the study of quantum
 properties of D-branes coupled to the closed string background fields 
$g_{\mu\nu}$ and $B_{\mu\nu}$. The boundary conditions of open strings 
of D-branes interpolate between Neumann and Dirichlet conditions and depend
 on the value of the  NS-NS $B_{\mu\nu}$ field \cite{k,l}. For large values of B (or equivalently  B constant if one uses the Seiberg-Witten limit taking the closed string metric $g_{\mu\nu}$ to zero) the action of the system is 
dominated by its boundary terms describing a  boundary world-sheet conformal
 field theory. In this case the correlation functions of the one dimensional
 boundary string fields $x^M(t)$ satisfy nontrivial commutation relations leading to a noncommutative space-time. This noncommutative space  has a
 canonical geometry extending  the usual phase space geometry of quantum 
mechanics \cite{m,n}. The coordinates $x^M(t)$ do no longer commute and the 
usual product of functions is  replaced by the  star product of Moyal 
Bracket \cite{o}. Soon after this development several studies have been 
devoted to develop the quantum field theory on noncommutative spaces \cite{p,q} and many partial results have been obtained \cite{r,s,t}.

The aim of this paper is to contribute to these efforts  
by considering the problem of noncommutative instantons in harmonic space
 (NHS). Our main motivations for this are: (1) harmonic space is a natural
 space where the problem of solving the self-dual Yang-Mills constraints may
 be done in straightforward way due to the important role played by harmonic
 analyticity discovered by Galperin {\it  et al} in the mid-eighties. 
This latter has known many successful applications as in the off-shell
 formulation of extended supersymmetric and supergravity theories \cite{u,v,w}, 
in hyperKahler metrics building \cite{x,y,z} and in the study of Yang-Mills 
self-dual solutions \cite{ab,ac}. (2) Standard noncommutative instantons analysis shows that the self-dual constraint eqs are non-linear and hence  difficult to solve exactly  \cite{a}.  The known non trivial solutions are obtained by perturbative analysis around the ordinary geometry. But here we will develop a different method based on a noncommutative harmonic space  leading to non perturbative explicit solutions. We will see that NHS method 
 offers a powerful manner to go beyond the standard perturbative solutions. 

The presentation of this paper is as follows. In section 2, we review the 
main lines of the problem of solving self-dual Yang-Mills constraints eqs 
in ordinary HS. In section 3, we study the Yang-Mills instantons on NHS.  
We first construct the NHS spaces and show the existence of two subspaces
 NHS($\eta$,0) and  NHS(0,$\theta$) depending on the values of the  
deformation parameters $\theta$  and $\eta$. Then we establish the connection between these noncommutative geometries and superstrings theory in presence of background fields $g_{\mu\nu}$ and $B_{\mu\nu}$. Next we focus our attention on NHS($\eta$,0) and study the perturbative solutions of the self-dual constraint eqs  while the derivation of the  exact solutions of these constraints is given in section 4. Finally, section 5 will be devoted to the conclusion.   

\section{Self-Dual Yang-Mills Constraints}

\hspace{1cm} As promised in the introduction, we review here briefly the main lines of one of possible resolution ways of the problem of finding self-dual Yang-Mills (YM) solutions in ordinary ${\R}^4$. The method we will present is a powerful way exhibiting manifestly the SU(2) symmetry rotating the three Kahler structures of the hyperKahler moduli space of instantons. It allows in addition to reformulate the self-dual YM constraint eqs as integrability conditions for the existence of analytic homeomorphisms of patches of ${\bf C}^2\sim{\R}^4$. This method permits also to construct explicit solutions of the self-dual constraints by working in harmonic space based on first realising geometrically the  SU(2)$\simeq{\bf S}^3$ symmetry 
of the instanton hyperKahler moduli space and second using the 
concept of harmonic analyticity extending the usual complex analyticity where only one Kahler structure is manifestly exhibited.

To fix the ideas, we will start by describing the fundamentals of harmonic
 analyticity for YM instantons. Recall that such analyticity was successfully
exploited in different occasions. In particular in the off-shell superspace
 formulation of supersymmetric field theories with eight supercharges \cite{u,ad,x}, $d=4$ ${\cal N}=2$ and $d=2$ ${\cal N}=4$ (conformal) supergravity theories \cite{w,ae,af,ag}  and in the hyperKahler metrics building \cite{x,y,z,ah}.  

\subsection{Harmonic Analyticity}

\hspace{1cm}Since the analysis we will present hereafter 
is valid for generalized self-dual YM fields in ordinary
 ${\R}^{4n}$, $n=1,2,\ldots $, we shall give the machinery for 
${\R}^{4n}$ and consider the particular leading case whenever it is necessary.
 Roughly speaking  ${\R}^{4n}$ is a real commutative Euclidean flat space 
whose local coordinates $\{x^M ,\;M=1,\ldots,4n\}$ obey generally the 
following natural identities
\bea\label{rch}
&{\overline{x^M}}&=x^M\nn\\
&x^Mx^N-x^Nx^M&=0\\
&x^M \to x'^M&=\Lambda ^M_N x^N,\nn
\eea
which define the reality, commutativity and homogeneity conditions
 respectively. The $\Lambda^M_N$'s are SO($4n$) Lorentz matrices in 
the ${\bf 4n}$ vector representation. If ${\R}^{4n}$ is endowed by a
 complex structure, the natural coordinates are the usual complex 
variables $z^{\a}$ and of $z_{\bar {\a}}$ of ${\bf C}^{2n}$. They
 transform in the fundamental ${\bf{2n}}$  and ${\bf{\overline{2n}}}$
 representations of U($2n$)=U(1)$\times$SU($2n$) $\subset$ SO($4n$) and 
related to the $x^M$'s as
\bea\label{ceq}
z^{\a}&=&x^{\a}+ix^{\a+2n};\nn\\
z_{\bar{\a}}&=&x^{\a}-ix^{\a+2n} ,\;\;\; \a=1,\ldots,2n.
\eea
Reality, commutativity and homogeneity of ${\R}^{4n}$ then become in 
${\bf C}^{2n}$
\bea\label{crch}
\[z^{\a},z^{\b}\]=[z_{\bar{\a}},z_{\bar{\b}}]=[z^{\a},z_{\bar{\b}}]=0\nn\\
 z^{\a}=u^{\a}_{.{\b}}z^{\b},\;\;\;z'_{\bar{\a}}=z_{\bar{\b}}u^{\bar{\b}}_{.\bar{\a}}
\eea
where $u^{ \bar{\a}}_{.\bar{\b}}\in{\bf{2n}}$ and $u^{\bar{\b}}_
{.\bar {\a}}\in{\bf{\bar{2n}}}$ with the property $z_{\bar{\a}}=(z^{\a})^+$.
 Though this complex frame exhibits manifestly one complex structure, one can
 go ahead and study the problem of solving the YM self-duality constraint eqs
 on ${\bf C}^2$. One recalls for instance the ADHM construction of YM 
instantons which find actually many applications in D-brane physics.
 Using this complex frame one can do even more by considering the problem 
of YM fields on ${\bf C}^{2n},\;n\geq 1$ and study the solving of the 
generalized self-dual constraint eqs. Such a kind of problem was considered 
in many occasions before as in YM and gravitational theories on ${\bf C}^p$ 
in connection with the Yang-Lee theory and (hyper)Kahler geometry respectively \cite{ai,aj,ab,ak}. In the present study, we shall not use the complex frame as given in 
(\ref{ceq},\ref{crch}). What we will do instead is to use the SU(2) symmetry 
factor of the SU(2)$\times$ SP($n$) homogeneity group
of $n$ dimensional hyperKahler manifolds to introduce a new local frame
 of ${\bf C}^{2n}$ where the three complex structures are manifestly 
exhibited. Breaking the SO($4n$) Lorentz group of ${\R}^{4n}$ down to 
SU(2)$\times$ SP($n$) by reindexing the $x^M$ variables as $x^{i\a}$,
 where now the double index $(i,\a)$ transform in the $({\bf 2},{\bf 2n})$ 
representation of SU(2)$\times$ SP($n$), 
eqs (\ref{rch}) read then as
\bea\label{rcco}
&{\bar{x}}^{i\a}&=\Omega_{\a\b}{\varepsilon_{ij}}x^{j\b},\nn\\
&\[x^{i\a},x^{j\b}\]&=0\\
&x^{i\a}\to x^{i\a'}&= u^i_j C^{\a}_{\b} x^{j\b},\nn
\eea
where $u^i_j$ and $C^{\a}_{\b}$ are respectively $2\times2$ and $2n\times2n$ 
matrices in the fundamental representations of SU(2) and SP($n$), while
 $\Omega$ and $\varepsilon$ satisfy the following
properties $\Omega^{\a\G}\Omega_{\G\delta}=\delta^{\G}_{\delta},\;\;\Omega^{
\a\b}=-\Omega^{\b\a},\;\Omega^{\a,\a +n}=1 ; 1 \leq\a\leq n$ and $\varepsilon^{ik}\varepsilon_{kj}=\delta^i_j,\;
\varepsilon^{ij}=-\varepsilon^{ji},\;\varepsilon^{12}=\varepsilon_{21}
=1 $. \\
To write down the generalized self-dual YM constraint eqs using the local
 coordinates system $\{x^{i\a}\}$, it is enough to consider the gauge
 covariant derivatives ${\cal D}_{i\a}$ and the gauge curvatures 
$F_{i\a j\b}$. Like for YM theory on ${\R}^{4n}$ with gauge group {\bf G},
 we have
\bea
{\cal D}_{i\a}&=&{\p\o{\p x^{i\a}}}+A_{i\a}\nn\\
&=&\p_{i\a}+A_{i\a},
\eea
where  $ A_{i \a}= A_{i \a}(x^{j \b})$ are the gauge potential component
 fields valued in the Lie algebra ${\bf g}$ of the gauge group {\bf G}. 
The field strengths $F_{i\a j\b}$ are given by the commutators of the  
$ {\cal D}_{i \a}$'s
\be
[{\cal D}_{i\a},{\cal D}_{j\b}]=iF_{i\a j\b}.
\ee
Taking the tensor product of the fundamental representation of $SU(2)\times SP(n)$, one can put the above eqs into a convenient form by decomposing $F_{i\a j\b}$  curvatures as shown here below
\be\label{derv}
[{\cal D}_{i\a},{\cal D}_{j\b}]=i \varepsilon_{ij} F_{(\a\b)}+ iF_ {(ij)[\a\b]},
\ee
where ( ) and [ ] denote complete symmetrisation and antisymmetrisation 
respectively. Having given the curvatures $F_{i\a j\b}$, we turn now to 
introduce the generalized self-duality constraint eqs of YM fields on
 $4n$ dimensional space with SU(2)$\times$ SP($n$) homogeneity symmetry.
 They are defined as
\be\label{sdc}
2iF_ {(ij)[\a\b]}=([{\cal D}_{i\a},{\cal D}_{j\b}]+[{\cal D}_{j\a},{\cal D}_{i\b}])=0,
\ee
or equivalently
\be\label{sdce}
[{\cal D}_{i\a},{\cal D}_{j\b}]= i \varepsilon _{ij}F_ {(\a\b)}.
\ee
Note that for $n=1$, the homogeneity group of ${\R}^4$ is
 SU(2)$\times$ SP(1)$\sim$ SO(4). In this case, $ F_{(ij)[\a\b]}=\Omega
_{\a\b}F_{(ij)}$ and so eq (\ref{derv}) becomes 
 \be
[{\cal D}_{i\a},{\cal D}_{j\b}]=i\varepsilon _{(ij)}F_ {(\a \b)}+i\Omega _{\a\b}F_{(ij)}.
\ee
In this  special situation one may impose not only the self-dual constraint eqs, 
$ F_{(ij)}=0$, but also anti self-dual ones associated with the condition 
$ F_{(\a\b)}=0$.

What we want to do now is to show that in some special situations one may
 solve explicitly the generalized self-dual constraints (\ref{sdc},\ref{sdce}) by using harmonic analyticity. The method may be summarized as follows: First,
 instead of using the local coordinates systems  $\{x^{i\a}\}$  
parameterizing the $4n$ dimensional homogeneous space  ${\cal P}/SU(2)
\times SP(n)$, where ${\cal P}$ stands for the Poincar\'e group with SU(2)
$\times$  SP($n$) as its Lorentz subgroup, we rather use the harmonic space   ${\cal P}$/SP($n$) parametrized by $\{x^{\pm\a},u^{\pm}_i\}$ . 

On this 
new space we have the following:\\
(i) the $u^{\pm}_i$ variables parameterize the unit ${\bf S}^3$ sphere. In terms of these variables, the  SU(2) Lorentz sub-symmetry of the $4n$ space is generated by the differential operators 
\bea\label{deq}
&\D &=u^+\p/\p u^-\nn\\
&\d&= u^-\p /\p u^+\\
&D^{0}&=u^+\p/\p u^+-u^-\p /\p u^-,\nn
\eea
which obey the familiar su(2) algebra
\bea\label{dceq}
&\[\D,\d\]&=D^{0}\nn\\
&\[D^{0},\D\]&=2\D\\
&\[D^{0},\d\]&=-2D^{0}\nn
\eea
(ii) the ${\X}^{\a}$ and ${\x}^{\a}$  bosonic coordinates still carry 
 an su(2) isospin $1\o 2$   representation since
\bea\label{salma}
&\[\D,{\X}^{\a}\]&=0\nn\\
&\[\D,{\x}^{\a}\]&={\X}^{\a}\\
&\[D^{0},{x^{\pm}}^{\a}\]&=\pm{x^{\pm}}^{\a}.\nn
\eea
 Eqs (\ref{salma}) mean that ${\X}^{\a}=u^+_ix^{i\a}$ and ${\x}^{\a}=u^-_ix^{i\a}$. Note that, as far as the $x^\pm$'s are concerned, the usual reality condition $ {\wt{x^{i\a}}}= \varepsilon _{ij}\Omega_{\a\b}x^{j\a}$
 is replaced in HS by 
\bea
{\wt{{\X}^{\a}}}&=&-\Omega_{\a\b}{\X}^{\b}\nn\\
{\wt{{\x}^{\a}}}&=&\Omega_{\a\b}{\x}^{\b}.
\eea
This conjugaison has been introduced first in \cite{u} and turns out to be the natural conjugation in HS.\\
(iii) The gauge covariant derivatives in HS language are defined as
\bea
{\cal D}^+_\a= \p/\p {\x}^{\a}+iA^+_\a=\p^+_\a +iA^+_\a\nn\\
{\cal D}^-_\a= \p/\p {\X}^{\a}+iA^-_\a=\p^-_\a +iA^-_\a.
\eea
They obey the following constraint eqs
\bea\label{dssc}
&\[\D,{\cal D}^+_\a\]&=0\nn\\
&\[\d,{\cal D}^+_\a\]&={\cal D}^-_\a\\
&\[D^{0},{\cal D}^{\pm}_\a\]&=\pm{\cal D}^{\pm}\nn
\eea
which ensure that ${\cal D}^{\pm}_\a$ are SU(2)  doublets; i.e. ${\cal D}^{\pm}_\a= u^{\pm}_i {\cal D}^{i}_\a$.\\
(iv) In HS, the gauge curvatures read as
\bea\label{gaugcur}
\[{\cal D}^{+}_{\a},{\cal D}^{+}_{\b}\]&=&i F^{++}_{[\a\b]}\nn\\
\[{\cal D}^{+}_{\a},{\cal D}^{-}_{\b}\]&=&i F_{(\a\b)}+iF^{+-}_{[\a\b]}
\eea
They verify the constraint eqs 
\bea
&\[{\D}^2,F^{+-}_{[\a\b]}\]&=\[{\cal D}^{++},F^{++}_{[\a\b]}\]=0 \nn\\
&\[{\D},F_{(\a\b)}\]&=0.
\eea
Eqs (\ref{gaugcur}) mean amongst others that the symmetric part $F_{(\a\b)}$
 of the curvature is a SU(2) singlet as it does not depend on the 
$u^{\pm}$'s while the antisymmetric part is a SU(2) triplet, i.e. 
\be
F^{++}_{[\a\b]}=u^+_{(i}u^+_{j)}F^{(ij)}_{[\a\b]}.
\ee
The second step consists of writing the generalized self-duality conditions 
(\ref{sdc},\ref{sdce}) in the HS as
\be
[{\cal D}^{+}_{\a},{\cal D}^{+}_{\b}]=0.
\ee
These constraints, which take a simple form in HS, can be treated as 
integrability conditions for the existence of an analytic harmonic subspace 
(AHSS) of the HS space. On this subspace, the solving of $F^{++}_{[\a\b]}=0$ 
may be easily worked out. The key idea of the AHSS resolution method is to say that under a gauge transformation{\footnote{This transformation is given in \cite{ab}}, one may set $ A^+_\a=0$ so that the gauge covariant derivatives $ D^+_\a$ reduces to a 
flat space derivative $ \p^+ =\p/\p{\x}^{\a}$. The price one should pay for 
this operation is that the su(2) harmonic derivatives $\D$, $\d$ and $D^{0}$
 eqs (\ref{deq}) acquire now connections as shown here below 
\bea\label{hdg}
&\D\to{\cal D}^{++}&=\D+i\V\\\nn
&\d\to {\cal D}^{--}&=\d+i\v\\
&D^{0}\to {\cal D}^{0}&=D^{0},\nn
\eea
with the requirement
\bea\label{gdsu}
&\[{\cal D}^{++},{\cal D}^{--}\]&=D^0\nn\\
&\[D^{0},{\cal D}^{\pm\pm}\]&=\pm 2{\cal D}^{\pm\pm},
\eea
or equivalently
\bea\label{vmvp}
&\[D^0 ,\V\] &=2\V, \nn\\
&\[D^0 ,\v\] &=-2\v.\\
&\D\v-\d\V+i\[ V^{++},V^{--}\] &= 0, \nn
\eea

Before going ahead, let us make some remarks regarding this tricky approach. 
$\V$ and $\v$  are harmonic gauge connections; they are not independent as 
they are related through eqs (\ref{vmvp}). Up on solving them, one may usually express $\v$ as a function of $\V$.  We will show later on how this is done in practice. For the moment let us note that $ \V$ turns out to
 be the basic object in solving the generalized self-duality constraint eqs.
 This property may be easily seen by rewriting the full set of the 
generalized self-duality constraints in HS. We have:
\bea\label{mgdsv}
&\[D^+_{\a}, D^+_{\b}\] &=0 \nn\\
&\[D^+_{\a},D^-_{\b}\] &=iF_{(\a\b)}\\
&\[D^0 , D^{\pm}\] &=D^ \pm_{\a},\nn
\eea
and
\bea\label{mgdsg}
&\[{\cal D}^{++} , D^+_{\a}\] &= 0\nn\\
&\[{\cal D}^{--} ,D^+_{\a}\] &=D^-_{\a}\nn\\
&\[{\cal D}^{++} ,{\cal D}^{--}\] &=D^0 \nn\\
&\[{\cal D}^0 ,{\cal D}^{\pm\pm}\] &=\pm 2{\cal D}^{\pm\pm}.
\eea
To derive the solutions of these constraints eqs, let us rewrite them in 
a more convenient way. Using harmonic analyticity which allows to take 
${\cal D}^+_{\a}$ as $\p /\p {\x}^{\a}$; i.e.  $ A^+_{\a}=0 $, and replacing 
 ${\cal D}^{++}$, ${\cal D}^{--}$ and ${\cal D}^{0}$ by their expressions 
(\ref{hdg}), the generalized self-duality constraint eqs may be represented 
as
\bea\label{adil}
&\[D^0, \V\] &=2\V \qquad (a)\nn\\
&\[\p^+_{\a} ,\V\] &= 0\qquad\qquad(b)\nn\\
&\D\v- \d\V+i\[\V,\v\]&=0\qquad\qquad (c)\\
&\[\p^+_{\a},\v\]&=-A^-_{\a}\;\;\qquad (d)\nn\\
&\[\p^+_{\a},A^-_{\b}\]&=F_{(\a\b)}.\;\qquad (e)\nn
\eea
These relations define the form of the self-duality YM constraint eqs we are looking for.

\subsection {Solving The Constraints}
\hspace{1cm}To solve the previous eqs, we shall proceed step by step by working out the solution of each relation of the system (\ref{adil}). These details  will be useful in the study of the instantons in NHS.  The first relation (\ref{adil}.a) tells us that  $\V$ is a harmonic function on the  
${\bf S}^2\simeq SU(2)/U(1)$ sphere and so it may be expanded in a harmonic
 series in term of the $u^{\pm}$'s as:
\bea\label{vsol}
\V(\X,\x,u^\pm)&=&\V(\X,\x)+ u^+_{(i}u^-_{j)}{\V}^{(ij)}(\X,\x)\nn\\
&&+u^{+}_{(i_1}u^{+}_{i_2}u^{-}_{j_1}u^{-}_{j_2)}{\V}^{(i_1i_2j_1j_2)}(\X,\x)+\ldots,
\eea
where the fields ${\V}^{(i_1i_2\ldots i_n,j_1j_2\dots j_n)}(\X,\x)$ are 
coefficients of the harmonic expansion on ${\bf S}^2$. It turns out that
 except the leading term $\V(\X,\x)$ where no explicit dependence on the 
$u^{\pm}$'s  appear, all the extra terms of (\ref{vsol}) may be ignored due to the existence of U(1) symmetry see (\ref{rachid}). Forgetting about these terms, and putting (\ref{vsol}) in (\ref{adil}.b), one sees that the potential  $\V$ should be an analytic function on ${\bf S}^2$ depending on the $\X$ variable only. This means that the general solution of (\ref{adil}) for a gauge group {\bf G} of generators $\{T^a\}$ should read as
\bea
(\V)^a&=&{\rm {tr}}(T^a\V) \nn\\
&=&{\cal M}^a_{\a\b}{\X}^{\a}{\X}^{\b},
\eea
where ${\cal M}^a_{\a\b}$  are constant coefficients scaling as the inverse
 of the square length dimension. Note that ${\cal M}^a_{\a\b}$ carry two 
SP($n$) indices $ \a,\b$. Commutativity of ${\X}^{\a}$ and ${\X}^{\b}$
 requires ${\cal M}^a_{\a\b}= {\cal M}^a_{\b\a}$ and so transform under the 
adjoint representation of the SP($n$) Lorentz subgroup. Note also that in the
 case where the gauge group {\bf G} is itself a SP($n$) symmetry, one may 
take as a solution for the SP($n$) potentials ${\V}_{\a\b}$ the following 
remarkable one
\be\label{pvsol}
(\V)_{\a\b}={i\o{\rho}^2}\Omega_{\a\G}\Omega_{\b\delta}{\X}^{\G}{\X}^{\delta}
\ee
where $\rho$ is a dimensionful parameter interpreted in  ${\R}^4$
as the size of the SU(2) instanton. Putting this solution back into the 
constraint (\ref{adil}.c), one gets after some straightforward algebra the 
following expression for the SP($n$) gauge potentials: 
\be\label{mvsol}
(\v)_{\a\b}={i\o{ (x^2+\rho^2)}}\Omega_{\a\G}\Omega_{\b\delta}{\x}^{\G}{\x}^{\delta},
\ee
where we have used the convention notation $x^2=\Omega_{\a\b} 
{\X}^{\a}{\x}^{\b}={\X}^{\a}{\x}_{\a}$. From (\ref{mvsol}), one gets the gauge 
potential $A^-_{\a}$ by help of (\ref{adil}.d). It reads as:
\be
(A^-_{\nu})_{\a\b}={-i\o{(x^2+\rho^2)}}\[-\Omega_{\nu(\a}\Omega_{\b)\G}{\x}^{\G}+ {1\o{(x^2+\rho^2)}}\Omega_{\nu\sigma}\Omega_{\a\G}\Omega_{\b\delta} {\X}^{\sigma}{\x}^{\G}{\x}^{\delta}\]
\ee
The curvatures $(F_{\mu\nu})_{\a\b}$ are immediately obtained by solving
 (\ref{adil}.e) using the relation of $A^-_{\a}$ given here above. We find:
\bea\label{fsol}
(F_{\mu\nu})_{\a\b}&=&{1\o{(x^2+\rho^2)}}\[-\Omega_{\mu(\a}\Omega_{\b)\nu}+ {1\o{(x^2+\rho^2)}}\Omega_{\mu(\a}\Omega_{\b)\G}\Omega_{\nu\sigma} {\X}^{\sigma}{\x}^{\G}\]\nn\\
&&-{i\o{(x^2+\rho^2)^2}}\Omega_{\mu\rho}\Omega_{\nu(\a}\Omega_{\b)\G}{\X}^{\rho}{\x}^{\G}\nn\\
&&{2i\o{(x^2+\rho^2)^3}}\Omega_{\mu\rho}\Omega_{\nu\sigma}\Omega_{\a\G}\Omega_{\b\delta}{\X}^{\rho}{\X}^{\sigma}{\x}^{\G}{\x}^{\delta}
\eea
Finally the Lagrangian density of the instanton is:
\be\label{sahraoui}
{\rm {tr}}\[(F_{\mu\nu})^2\]={{\rho^4}\o{(x^2+\rho^2)^4}}.
\ee
In the limit $\rho\to 0$, eq (\ref{sahraoui}) has a singularity near the origin 
and so the density tr$(F_{\mu\nu})^2$ behaves as a Dirac delta function.

\section{Yang-Mills Instantons on Noncommutative ${\R}^4_\theta$} 
\hspace{1cm}We start this section by considering the extension of the harmonic space analyticity idea in order to study the problem of Y-M instantons on noncommutative ${\R}^4_\theta$. To do so, we have a priori different choices of non commutative local coordinates. For example one can use directly the usual real coordinates system $\{x^M\}$, where the $x^M$'s 
 transforming in the {\bf 4} vector representation of SO(4) satisfy:
\be
 \[x^M,x^N\]=i\theta^{MN}.
\ee
Then extends the standard analysis of YM instantons by incorporating this 
 noncommutativity feature. An other way is to break the SO(4) Lorentz group 
down to U(2)$=$U(1)$\times$SU(2) and use the complex coordinates frame  
 $\{z^{\mu}=x^{\mu}+ix^{\mu+2};\; iz^{\bar{\mu}}=x^{\mu}-ix^{\mu+2};\;
 \mu=1,2\}$ satisfying the following commutation relations
\bea\label{ncc}
\[z^\mu,z^\nu\]&=&\theta^{\mu\nu}\nn\\
\[z^{\bar{\mu}},z^{\bar{\nu}}\]&=&\theta^{\bar{\mu}\bar{\nu}}\nn\\
\[z^\mu,z^{\bar{\nu}}\]&=&\theta^{\mu\bar{\nu}}\nn\\
\[z^{\bar{\mu}},z^\nu\]&=&\theta^{\bar{\mu}\nu}
\eea
This coordinate system, which is useful whenever one can impose complex
analyticity, is nowadays intensively used in the noncommutative ADHM 
formulation of instantons \cite{e,f,g}. But we will use neither the first method nor the second. What we will do instead is to develop a non commutative harmonic space analysis by extending the study of subsection 2.2. 
Thus taking the local coordinates of ${\R}^4$ as  $\{x^{i\a};\;\;i=1,2;\; 
\a=1,2\}$. The space of noncommutativity reads as
\be\label{nch}
 \[x^{i\a},x^{j\b}\]= \Omega^{\a\b}\eta^{(ij)}+\varepsilon^{ij}\theta^{(\a\b)},
\ee
where $\eta^{(ij)}$ and $\theta^{(\a\b)}$ transform respectively under the
 $(1,0)$ and $(0,1)$ representations of the SU(2)$\times$SP(1) Lorentz group.
 Now introducing the harmonic variables $u^{\pm}_i$, the noncommutative 
harmonic space is defined as
\bea\label{nchs}
\[{\X}^{\a},{\X}^{\b}\]&=&i\Omega^{\a\b}\pe\nn\\
\[{\x}^{\a},{\x}^{\b}\]&=&i\Omega^{\a\b}\me\nn\\
\[{\X}^{\a},{\x}^{\b}\]&=&i\Omega^{\a\b}\eta-i\theta^{(\a\b)}\nn\\
\[{\x}^{\a},{\X}^{\b}\]&=&i\Omega^{\a\b}\eta+i\theta^{(\a\b)},
\eea
where we have used
\bea
&\pe&= u^+_{(i}u^+_{j)} \theta^{(ij)}\nn\\
&\me&= u^-_{(i}u^-_{j)} \theta^{(ij)}\\
&\eta&= u^+_{(i}u^-_{j)} \theta^{(ij)}.\nn
\eea
Observe that according to the values of $\eta^{(ij)}$ and $\theta ^{(\a\b)}$ 
which together carry the six degrees of freedom of the six-dimensional 
antisymmetric representation of SO(4), one may distinguish four kinds of  
harmonic subspaces: (1) the ordinary one considered in section 2 
corresponding to $\theta^{(ij)}=0$; $\eta^{(ij)}=0$  and three noncommutative
 ones given by (2) $\eta^{(ij)}\neq 0$; $\theta ^{(\a\b )}=0$, 
 (3) $\eta^{(ij)}=0$; $\theta^{(ij)}\neq 0$ and (4) $\eta^{(ij)}\neq 0$; 
 $\theta ^{(\a\b)}\neq 0$. We shall refer to these four kinds of harmonic 
spaces as HS($\eta,\theta$) which roughly speaking may be viewed as a two
 deformation parameters of the standard harmonic space. Note moreover that 
these spaces are intimately related with the problem of finding (anti)
 self-dual YM instantons on noncommutative geometry. In this regards it is not
 difficult to see that the noncommutative harmonic subspaces 
HS($\eta$,0)  and HS(0,$\theta$) are respectively associated with self-dual 
and anti self-dual YM instantons of the noncommutative geometry. The point is
 to use the Seiberg-Witten realisation according to which in the $\a'$ zero slope limit the parameters $\eta^{(ij)}$ and $\theta^{(\a\b)}$ are proportional to the inverse of the NS-NS antisymmetric $B_{MN}$ field leaving on an Euclidean D3-Brane world volume. Put differently, if we decompose
 $B_{i\a j\b}$ as $\varepsilon_{ij}B_{(\a\b)}+\Omega_{\a\b}B_{(ij)}$; we have
\be
\eta^{(ij)}\sim B_{(ij)}/B^2;\;\;\;\; \theta^{(\a\b)}\sim B_{(\a\b)}/B^2.
\ee
These eqs show clearly that $\eta^{(ij)}$ is related to the self-dual
 part of the $B$ field while $\theta^{(\a\b)}$ is given by the 
anti self-dual one. In what follows we shall consider the example of
 noncommutative HS($\eta$,0) and study the problem of finding 
noncommutative YM instantons by using the idea of harmonic analyticity
 developed previously. Similar analysis may be done for the 
anti self-dual constraints by using NHS(0, $\theta$).

\subsection {Noncommutative YM instantons in NHS($\eta$,0)}

\hspace{1cm}The commutation relations of the local coordinates $\{{\X}^{\a}, {\x}^{\b}, u^\pm_i\}$ defining the noncommutative HS($\eta$,0) are immediately obtained from (\ref{nchs}) by setting $\theta^{(\a\b)}=0$ which gives
\bea
\[{\X}^{\a},{\X}^{\b}\]&=&i\Omega^{\a\b}\pe \nn\\
\[{\X}^{\a},{\x}^{\b}\]&=&i\Omega^{\a\b}\eta \\
\[{\x}^{\a},{\x}^{\b}\]&=&i\Omega^{\a\b}\me \nn.
\eea
Such commutation relations have however a nice interpretation in D-brane physics in presence of closed string background fields 
$g_{i\a j\b}$ and $B_{i\a j\b}$. Indeed if we consider a D3-brane with
 Euclidean ${\R}^4$ world volume in a constant antisymmetric $B$ field
 and world sheet action
\be\label{wsa}
{\cS} ={1\o {4\pi\a'}}\int_{\S}\(g_{i\a j\b}\delta^{AB}-2i\pi B_{i\a j\b}\varepsilon^{AB}\)\p_{A}x^{i\a}\p_{B}x^{j\b},
\ee
consistency requires that the open string fields $x^{i\a}$ should obey
 the following boundary conditions
\be\label{bc}
\(g_{i\a j\b}\p_{n}x^{j\b}+2i\a' B_{i\a j\b}\p_{t}x^{j\b}\)|_{\p\S}=0,
\ee
where $\p_{n}$ and $\p_{t}$ are the normal and tangential derivatives along the world sheet boundary $\p\S$ respectively. In practice we will use the classical approximation to open string where $\S$ may be taken as a disc and hence can be conformally mapped to the upper half plane parametrized by 
$z=\tau +i\sigma$ and $\p={1\o 2}({\p\o {\p\tau}}-i{\p\o {\p\sigma}})$,0
 $\sigma \ge 0$. In this world sheet
 coordinates, the boundary conditions (\ref{bc}) becomes
\be\label{bcd}
\[g_{i\a j\b}\(\p-\bar{\p}\)x^{j\b}+2i\a' B_{i\a j\b}(\p+\bar{\p}\)x^{j\b}\]|_{z=\bar z}=0.
\ee
Eq (\ref{bcd}) describes an interpolation from Neumann to Dirichlet 
boundary conditions. In the Seiberg Witten limit where 
$g_{i\a j\b}\sim{\a'}^2\varepsilon_{ij}\Omega_{\a\b}$ with $\a'\sim 
\epsilon ^{1\o 2}$ goes to zero, Eq (\ref{bcd}) becomes Dirichlet boundary 
conditions where at each boundary the open string world sheet is 
attached to a single point (zero brane) in the D3 brane. In this limit
, the action (\ref{wsa}) reduces to
\be
{\cS} =-{i\o {2}}\int_{\R}d\tau \(B_{ij}\Omega^{\a\b}x^{i\a}{dx^{j\b}\o {d\tau}}+B_{\a\b}\varepsilon_{ij}x^{i\a}{dx^{j\b}\o {d\tau}}\).
\ee
Now taking the anti self-dual part $B_{(\a\b)}$ of the antisymmetric 
NS-NS field to zero, one can calculate the propagators of the boundary
 fields $x^{i\a}(\tau)$. From the eqs of motion of the one dimensional
fields $x^{k\G}(\tau)$, namely
\be
i\varepsilon^{ki}B_{(ij)}{dx^{j\G}\o {d\tau}}=0,
\ee
one can easily check that the propagators  of the $x^{i\a}(\tau)$'s read 
as 
\be
\langle x^{i\a}(\tau_1)x^{j\b}(\tau_2)\rangle ={i\o {2}}\eta^{(ij)}\Omega^{\a\b}\varepsilon(\tau_1-\tau_2),
\ee
where $\varepsilon(\tau_1-\tau_2)$ is the Heveaside function 
$\varepsilon(\tau)=1$ for $\tau >0$ and $\varepsilon(\tau)=-1$ for
 $\tau <0$ and where
\be
\eta^{(ij)}B_{(ij)}=1
\ee
In harmonic space $\{{\X}^{\a},{\x}^{\a}, u^{\pm}\}$, the boundary
 conformal field theory we have discussed is described by the 
following action
\be
{\cS} =-{i\o {2}}\int_{\R\times S^2}d\tau du
\Omega_{\a\b}\Bigl[\(B^{++}{\x}^{\a}{d{\x}^{\b}\o {d\tau}}-B^{--}{\X}^{\a}{d{\X}^{\b}\o {d\tau}}\)-B\( {\X}^{\a}{d{\x}^{\b}\o {d\tau}}+{\x}^{\a}{d{\X}^{\b}\o {d\tau}}\)\Bigr],
\ee
where integration with respect to the $u$'s keeps only SU(2) singlets and where we have used
\bea
&B^{++}&=u^+_{(i}u^+_{j)}B^{(ij)}\nn\\
&B^{--}&=u^-_{(i}u^-_{j)}B^{(ij)}\\
&B&=u^+_{(i}u^-_{j)}B^{(ij)}\nn.
\eea
The propagators in the harmonic space are
\bea\label{sdp}
\langle {\X}^{i\a}(\tau_1){\X}^{j\b}(\tau_2)\rangle &=&{i\o {2}}\Omega^{\a\b}\eta^{++}\varepsilon(\tau_{12}),\nn\\
\langle {\x}^{i\a}(\tau_1){\x}^{j\b}(\tau_2)\rangle &=&{i\o {2}}\Omega^{\a\b}\eta^{--}\varepsilon(\tau_{12}),\nn\\
\langle {\X}^{i\a}(\tau_1){\x}^{j\b}(\tau_2)\rangle &=&{i\o {2}}\Omega^{\a\b}\eta\varepsilon(\tau_{12}),\nn\\
\langle {\x}^{i\a}(\tau_1){\X}^{j\b}(\tau_2)\rangle &=&{i\o {2}}\Omega^{\a\b}\eta\varepsilon(\tau_{12}),
\eea
Computing the commutators of the conformal field operators
 ${x^{\pm}}^{\a}(\tau)$ by using the short distance products (\ref{sdp}),
 we find the following relations
\bea
\[{\X}^{i\a}(\tau_1),{\X}^{j\b}(\tau_2)\] &=&i\Omega^{\a\b}\eta^{++},\nn\\
\[{\x}^{i\a}(\tau_1),{\x}^{j\b}(\tau_2)\]&=&i\Omega^{\a\b}\eta^{--},\nn\\
\[{\X}^{i\a}(\tau_1),{\x}^{j\b}(\tau_2)\] &=&i\Omega^{\a\b}\eta,\nn\\
\[{\x}^{i\a}(\tau_1),{\X}^{j\b}(\tau_2)\] &=&i\Omega^{\a\b}\eta,
\eea
which are similar to the commutation relations of the noncommutative 
harmonic space NHS($\eta$,0) given by (\ref{nchs}). In the remainder of 
this section we want to discuss some features of the algebra of
 functions on NHS($\eta$,0). In conformal field theory on harmonic 
space one distinguishes different ground state vertex operators;
\be
\exp{(\pm ip^+\x)};\;\;\;\;\exp{(\pm ip^-\X)}.
\ee
They satisfy the short distance products
\bea
:e^{+ip^-\X}(1)::e^{+iq^-\X)}(2):&=&  e^{-{1\o 2}\pe p^-q^-}:e^{i(p^-+q^-)\X}(2):\nn\\
:e^{-ip^+\x}(1)::e^{-iq^+\x)}(2):&=&  e^{-{1\o 2}\me p^+q^+}:e^{i(p^++q^+)\x}(2):\nn\\
:e^{+ip^-\X}(1)::e^{-iq^+\x)}(2):&=&  e^{{1\o 2}\eta p^-q^+}:e^{+i(p^-\X -q^+\x)}(2):\nn\\
:e^{-ip^+\x}(1)::e^{-iq^-\X)}(2):&=&  e^{{1\o 2}\eta p^+q^-}:e^{-i(p^+\x-q^-\X)}(2):.
\eea
In noncommutative harmonic space language, the above short distance
 products coincide with the usual star product on noncommutative 
geometry. Thus we have for instance the identification
\be
\lim_{\tau\to 0}e^{\pm ip^-\X}(\tau).e^{\pm iq^-\X}(0)\sim e^{\pm ip^-\X}\* e^{\pm iq^-\X}.
\ee
More generally given two harmonic space functions $\f(\X,\x,u)$ and
 $\g(\Y,\y,u)$, the star product of these functions is defined as
\be\label{spfd}
\f (x)\*\g (y)=\exp {i\o 2}(\Omega^{\a\b}M_{\a\b})\f (x)\g (y),
\ee
where
\be
M_{\a\b}=\(\eta^{++}{\p\o {\p x^{+\a}}}{\p\o {\p y^{+\b}}}+
\eta^{--}{\p\o {\p x^{-\a}}}{\p\o {\p y^{-\b}}}\)-
\eta\( {\p\o {\p x^{+\a}}}{\p\o {\p y^{-\b}}}+{\p\o 
{\p x^{-\a}}}{\p\o {\p y^{+\b}}}\).
\ee
At first order in $\pe$, $\me$ and $\eta$, (\ref{spfd}) reduces to
\bea
\f\*\g &=&\f\g +{i\o 2}\Omega^{\a\b}\Big[\(\eta^{++}\p^-_{\a}\f\p^-_{\b}\g
+\eta^{--}\p^+_{\a}\f\p^+_{\b}\g\) \nn \\ 
&&-\eta\(\p^-_{\a}\f\p^+_{\b}\g+\p^+_{\a}\f\p^-_{\b}\g\)\Big]
+{\cal O}(2)
\eea
\subsection{ Self-Dual Yang-Mills Constraints in NHS($\eta$,0)}

\hspace{1cm}Noncommutative YM theory in Euclidean four dimensional space is 
formulated in a similar way as YM theory in ordinary ${\R}^4$
 except that the gauge group matrix multiplication is now replaced by 
the star product (\ref{spfd}). For instance the transformations of 
the gauge field ${\widehat A}_M$ and the field strength ${\widehat F}_
{MN}$ of noncommutative YM theory are   
\bea
\delta_{\ht}\ha_\mu &=&\p_{\mu}\ht -i\(\ht\*\ha_{\mu}-\ha_{\mu}\*\ht\) \nn \\
\hf_{MN}&=&\p_M\ha_N-\p_N\ha_M-i\(\ha_M\*\ha_N-\ha_N\*\ha_M \)\\
\delta_{\ht}\hf_{MN}&=&i\(\ht\*\hf_{MN}-\hf_{MN}\*\ht\). \nn
\eea
Remark that the gauge parameters $\ht$, the fields ${\widehat A}_M$
 and ${\widehat F}_{MN}$ carry a hat in order to be distinguished from
 their ordinary geometry analogue. This convention  notation will also
 be used in the remainder of this study.

 YM gauge theory in noncommutative harmonic space may be 
constructed in a similar manner as  in the standard formulation.
 This is achieved in practice by extending the usual HS(0,0) classical
 fields to functionals on NHS($\eta$,0) and replacing the ordinary 
product by the star one given by (\ref{spfd}). The novelty brought by 
the harmonic variables is completely controlled by the SU(2) symmetry. 
 Since the $u^{\pm}_i$ variables still obey the same relations as in
 ordinary HS(0,0), the covariant harmonic derivatives in NHS($\eta$,0)
 defined as
\bea\label{cdrn}
&\cd^{++}&=\D +i\hV \nn \\
&\cd^{--}&=\d +i\hv \\
&{\cal D}^{0}&=D^0, \nn 
\eea
where $\hV$ and $\hv$ are harmonic gauge connections on HS($\eta$,0), 
still obey the SU(2) algebra (\ref{gdsu}) which requires in turns:
\bea\label{crcdn}
&\[ D^0,\hV\]&=2\hV ; \nn \\
&\[ D^0,\hv\]&=-2\hv \\
&\D\hv -\d\hV +i\(\hV\*\hv-\hv\*\hV\)&=0. \nn
\eea
The SU(2) symmetry (\ref{gdsu}) and (\ref{cdrn},\ref{crcdn}) shows also
 that the self-dual YM 
constraint eqs in noncommutative NHS($\eta$,0) may be obtained from 
(\ref{adil}) by replacing the HS(0,0) objects by their extensions on
 NHS($\eta$,0). The noncommutative self-dual YM constraints read
 then:
\bea\label{mostapha}
&\[ D^0,\hV\]&=2\hV \qquad (a)\nn \\
&\[\p^+_{\a},\hV\]&=0 \qquad\qquad (b)\nn \\
&\D\hv -\d\hV &=-i\(\hV\*\hv -\hv\*\hV\)\qquad (c)\\
&\[\p^+_{\a},\hv\]&=-\ha^-_{\a} \nn \qquad(d) \\
&\[\p^+_{\a},\ha^-_{\b}\]&=\hf_{(\a\b)}\qquad (e) \nn
\eea
To solve these constraint eqs, we shall make two analysis by 
considering first perturbative solutions around the ordinary one.
 Then we examine the exact solution of these constraints by using 
noncommutative calculus on NHS($\eta$,0).\\

\subsection{Perturbative Analysis}

\hspace{1cm}Here we shall consider special perturbative solutions preserving 
manifestly the SU(2) symmetry and corresponding to small values of the
 deformation parameters $\pe$, $\me$ and $\eta$. Eqs (\ref{mostapha}) suggest
 that one may expand the noncommutative connections $\hV$ and $\hv$ 
around the ordinary $\V$ and $\v$  ones of (\ref{pvsol}, \ref{mvsol})
 as 
follows
\bea\label{fos}
\hV &=& \V +\eta\W +{\cal O}(2) \nn \\
\hv &=& \v +\eta\u +\me U +\pe U^{-4} +{\cal O}(2)
\eea
Expanding (\ref{mostapha}.c) to first order in $\pe$, $\me$ and $\eta$
 as in (\ref{spfd}) and using (\ref{mostapha},\ref{fos}), in particular
 the analyticity of $\hV$, we find
\bea
&\eta&\(\D\u +2U\)+\me\D U \nn\\
&&+\pe \(\D U^{-4}+\u\) 
-\eta\d\W -\me\W \nn \\
&&=-i\(\eta\{ \[\V,\U\]+\[\W,\v\]\}\nn\\
&&+\me\[\V,U]+\pe\[\V,U^{-4}\]\)
+{1\o 2}\Omega^{\a\b}\{\pe\(\p^-_{\a}\V\p^-_{\b}\v \nn\\
&&-\p^-_{\a}\v\p^-_{\b}\V\)
-\eta \(\p^-_{\a}\V\p^+_{\b}\v -\p^+_{\a}\v\p^-_{\b}\V\)\}+{\cal O}(2).
\eea
These eqs imply in turn
\bea
&\D U-\W +i\[\V,U\]=0\nn\\
&\D\u -\d\W +2U+i\[\V,\u\]+i\[\W,\v\]={1\o 2}\Omega^{\a\b}\{\p^-_{\a}\V\p^+_{\b}\v \nn\\
&-\p^+_{\a}\v\p^-_{\b}\V\)\nn \\
&\D U^{-4}+i\[\V,U^{-4}\]+\u ={1\o 2}\Omega^{\a\b}\{\p^-_{\a}\V\p^-_{\b}\v\nn\\
&-\p^-_{\a}\v\p^-_{\b}\V\)
\eea
To solve these eqs, we choose the fields $\W$, $U$, $\u$ and $U^{-4}$ 
of the form
\bea
&{\W}^{\a}_{\b}&=a {\X}^{\a}{\X}_{\b}\nn \\
&U^{\a}_{\b}&=b_1 {\X}^{\a}{\x}_{\b}+b_2 {\x}^{\a}{\X}_{\b}\\
&{\u}^{\a}_{\b}&=c {\x}^{\a}{\x}_{\b}\nn
\eea
where $a$, $b_1$, $b_2$  and $c$ are parameters to be determined.
Lengthy but straightforward calculations lead to:
\bea\label{sara}
 &a&={1\o {\rho^4}}\nn\\
 &b_1&=-{1\o{2(x^2 + {\rho}^2)^2}}\nn\\
 &b_2&=-{{2x^2+\rho^2}\o{2{\rho}^2(x^2 + {\rho}^2)^2}}\nn\\
 &c&=-{1\o{(x^2 + {\rho}^2)^2}} .
\eea

\section{ Exact Solution}

\hspace{1cm}Here we give the exact solution of the self-dual YM constraint 
(\ref{mostapha}) on noncommutative HS($\eta$,0). As for ordinary 
HS(0,0) harmonic space, (\ref{mostapha}.a-b) show that $\hV$ is a harmonic
 function on NHS($\eta$,0) depending on ${\X}^\a$ only. This means that
 according to (\ref{mostapha}.a-b), $\hV$ may be written as
\be\label{spdn}             
(\V)^{\a\b}={\X}^{\a}\* A\* {\X}^{\b}+C^{++}\Omega^{\a\b}
\ee
where $A$ and $\C$ are harmonic functions without any dependence on 
the $x^{\pm}$'s. The $A$ and $\C$   scale as the inverse of length 
squared. Therefore, they behave as the inverse of $\eta$ since 
according to (\ref{nchs})), one can check that we have the following 
identities
\bea\label{spen}
{\X}_{\G}\*{\X}^{\G}&=&\X\*\X=i\pe \nn\\
{\x}_{\G}\*{\x}^{\G}&=&\x\*\x=i\me \nn\\
{\X}_{\G}\*{\x}^{\G}&=&\X\*\x=i\eta +z^2 \nn\\
{\x}_{\G}\*{\X}^{\G}&=&\x\*\X=i\eta -z^2,
\eea
where we have used  $\Omega_{\a\b}\Omega^{\a\b}=2$ and set  
$z^2={1\o 2}(\X\*\x-\x\*\X)\geq 0$. Note that (\ref{spdn},\ref{spen})
 satisfy
\bea
\d(\V)^{\a\b}&=&\({\X}^{\a}\* A\*{\x}^{\b}+{\x}^{\a}\* A\*{\X}^{\b}\)\nn\\
&&+{\X}^{\a}\*(\d A)\*{\X}^{\b}+\(\d\C\)\Omega^{\a\b}.
\eea
and
\bea\label{spcb}
&\D\pe &=0\nn\\
&\D\eta&=\pe\\
&\D\me &=2\eta. \nn
\eea

The next step is to find $(\v)^{\a\b}$ by solving (\ref{mostapha}.c). 
Harmonic analysis on $NHS(\eta,0)$ shows that we should look for a 
solution of the form
\bea\label{spsn}
(\v)^{\a\b}&=&{\x}^{\a}\* B\*{\x}^{\b}+{\X}^{\a}\* \e\*{\x}^{\b}\nn\\
&&+{\x}^{\a}\* \k\*{\X}^{\b}+{\X}^{\a}\* G^{(-4)}\*{\X}^{\b}+\h\Omega^{\a\b}
\eea
where $B$, $\e$, $\k$, $G^{(-4)}$ and $\h$ are to be determined . 
Taking the harmonic derivative of (\ref{spsn}), we get
\bea\label{sphd}
\D(\v)^{\a\b}&=&{\X}^{\a}\* \[B+\D\e\]\*{\x}^{\b}+{\x}^{\a}\* \[B+\D\k\]\*{\X}^{\b}\nn\\
&&+{\X}^{\a}\*\[\e+\k+\D G^{(-4)}\]\*{\X}^{\b}\\
&&+{\x}^{\a}\* \[\D B\]\*{\x}^{\b}+\D\h\Omega^{\a\b}.\nn
\eea
Moreover using (\ref{spdn}) and (\ref{spsn}) we have
\bea\label{spfs}
i\({\V}^{\a}_{\G}\*{\v}^{\G}_{\b}&-&{\v}^{\a}_{\G}\*{\V}^{\G}_{\b}\)\nn \\
&=&{\X}^{\a}\*\[iA\*(\X\x)\*B +iA\*(\X\X)\*\e\]\*{\x}^{\b} \nn\\
&&+{\x}^{\a}\*\[-iB\*(\x\X)\*A -i\k\*(\X\X)\*A\]\*{\X}^{\b}\\
&&+{\X}^{\a}\*\[iA\*(\X\x)\*\k +iA\*(\X\X)\*G^{(-4)}\nn\\
&&-\e\*(\x\X)\*A-iG^{(-4)}\*(\X\X)\*A\]\*{\X}^{\b}.\nn
\eea
Putting back (\ref{spdn},\ref{spcb}),(\ref{spsn}) and (\ref{sphd}) in
 the constraint (\ref{mostapha}.c) 
one gets an equation transforming in the ${\bf{1\oplus 3}}$ representation 
of the SU(2) symmetry which up on projecting it along the different
 U(1) cartan subsymmetry directions, we get the following system of 
four equations
\bea\label{saidi}
&\Bigl[\D\e +iA\*(\X\X)\*\e +\[1+iA\*(\X\x)\]\*B-A\Bigr]&=0\;\;(a) \nn\\
&\Bigl[\D\k -i\k\*(\X\X)\*A +B\*\[1-iA\*(\x\X)\]\*A\]-A\Bigr]&=0\;\;(b)\nn \\
&\Bigl[\D G^{(-4)}+\e\*\(1-i(\x\X)\*A\)& \nn\\
&+\[1+iA\*(\X\x)\]\*\k-\d A\Bigr]&=0\;(c) \nn\\
&\D B&=0\;\;(d)\nn\\
&\D\h-\d\C&=0\;\;(e)\nn\\
\eea
 To solve these, it is interesting to note the following:
(1) the $\eta$  parameter is a {\bf C} number independent of the 
$x^{\pm}$'s and so commute with the star product. it scales as 
(length)$^{-2}$. $\me$ and $\pe$ are proportional to the norm of the 
 sp(1) isospinor ${\X}^{\a}$ and ${\x}^\a$ which are no longer zero 
due to noncommutativity. (2) In the limit $\eta\to 0$, $\X\x\ge 0$ and
 $\x\X\le 0$ as may be check explicitly by help of (\ref{rcco}) (3) Dimensional
 arguments show that the unknown functions  $A$, $B$, $\c$,
 $\e$,$\k$ and $G^{(-4)}$ in (\ref{saidi}) scale as
\bea
\[A\]&=&\[B\]=\[\e\]=\[\k\]=\[G^{-4}\]=-\[\eta\]\nn\\
&=& -2\[x^\pm\]=({\rm length})^{-2}
\eea
(4) In the limit $\eta\to 0$, eqs (\ref{saidi}) should coincide with 
the
 ordinary case and so  
\bea\label{limit}
\lim_{\eta\to 0}A&=-i&1\o{{\rho}^2}\nn\\
\lim_{\eta\to 0}B&=-i&1\o {\rho^2+z^2}\\
\lim_{\eta\to 0}\e&=&\lim_{\eta\to 0}\k=\lim_{\eta\to 0}G^{(-4)}=0.\nn
\eea
 These limits show that $\e$, $\k$ and $G^{-4}$  should be proportional
  to $\eta$. Eq (\ref{saidi}.d) shows that $B$ does
not depend on the harmonics. This means that if it carries a
 dependence on $\eta$, then this should be in terms of  the invariant
 $\pe\me-\eta^2 = \eta^{(ij)}\eta_{ij}={\vec{\eta}}^2$. However
 the dimensional  arguments show that the natural solution is just
 the  ordinary one which reads as 
\be\label{mounia}
B=-i{2\o{(2\rho^2+\X\x-\x\X)}}=-i{1\o{\rho^2+z^2}},
\ee
 where we have used  (\ref{spen}).

Now we turn to solve (\ref{saidi}.a-b) which we rewrite using 
(\ref{spen}) as 
\bea\label{sofab}
\[\D\e-\pe A\e\]+\[(1+iA\X\x)B-A\]&=&0\qquad (a)\nn\\
\[\D\k-\pe A\k\]+\[(1-iA\x\X)B-A\]&=&0\qquad (b).
\eea
Taking $A$ to be
\be\label{mouniab}
A=-i{1\o{\rho^2-i\eta}}
\ee
and using (\ref{mounia}), the last terms of eq (\ref{spen}.a) vanish. It follows 
that (\ref{sofab}) become
\bea
&\(\D +i{\pe\o{\rho^2-i\eta}}\)\e& =0\nn\\
&\(\D -i{\pe\o{\rho^2-i\eta}}\)\k+2\eta AB&=0.
\eea
 Integrating these eqs, we find that the most general solution is 
given by 
\bea\label{mgsigb}
\e&=&{\rm {cnst}}{{\rho^2-i\eta}\o{({\vec{\eta}}^2+\eta^2)}}\nn\\
\k&=&-\me AB \nn\\
&=&{{\me}\o{({\rho}^2-i\eta)(2\rho^2+\X\x-\x\X)}}
\eea
 However the  constant appearing in the solution  of (\ref{mgsigb}) 
should vanish due to the constraints (\ref{limit}). Therefore 
\be\label{mouniae}
\e=0
\ee
Now we turn to solve (\ref{saidi}.c). Using the previous solutions 
(\ref{mounia}), (\ref{mouniab}), (\ref{mgsigb}) and (\ref{mouniae}),
 eq  (\ref{saidi}.c) 
 can be brought to the  following form
\be
\D G^{(-4)}+{A\o B}\k=\d A.\nn
\ee
At first sight, this eq seems to be difficult to handle; but if 
we consider the following remarkable features
\bea
\k&=&-AB\me\nn\\
\d A&=&-\me A^2,
\eea
it reduces to
\be
\D G^{(-4)}=0
\ee
and so
\be
G^{(-4)}=0.
\ee
Finally, concerning  (\ref{saidi}.e), the $H^{--}$ and $\C$
 are solved as
\bea
\C&=&\D\lambda\nn\\
H^{--}&=&\d\lambda,
\eea
where $\lambda$ is an arbitrary function on the ${\bf S}^2$.

\bigskip\noindent
{\it Summary}

The noncommutative YM SU(2) instanton formulated on NHS($\eta$,0) is described by the harmonic connections ${\V}^{\a}_{\b}$ and ${\v}^{\a}_{\b}$ given
  \bea\label{hep}
{\V}^{\a}_{\b}&=&{\X}^\a\*A\*{\X}_\b+\D{\lambda}{\delta}^{\a}_{\b}\nn\\
{\v}^{\a}_{\b}&=&{\x}^\a\*B\*{\x}_\b-\me{\x}^\a\*A\*B\*{\X}_\b\\
&&+\d{\lambda}{\delta}^{\a}_{\b}\nn
\eea
with
\bea\label{phe}
A&=&-i{1\o{\rho^2-i\eta}}\nn\\
B&=&-i{1\o{\rho^2-i\eta+\X\x}}.
\eea
 At this level certain interesting comments  may be done: (1) The 
solutions of the noncommutative YM self-dual constraint eqs we
 have obtained are exact solutions. They extend the perturbative one given in the end of section 3.

(2) our solution admits a U(1) symmetry;
\bea\label{rachid}
\V&\rightarrow&\V+\D\lambda\nn\\
\v&\rightarrow&\v+\d\lambda
\eea
for any $\lambda$. (3) As long as $\eta\neq 0$, our solution is non 
singular. This result agrees with the absence of  small instanton in
noncommutative geometry. (4) In the limit $\eta \to 0$, we
recover the ordinary solution. Using the associativity of the star product,
(\ref{spfd}) and the algebra  of the differential operators
\bea
\nabla^-_{\mu}&=&\me\p^+_\mu -\eta\p^-_\mu  \nn\\
\nabla^+_{\mu}&=& \pe\p^-_\mu -\eta\p^+_\mu,
\eea
one discovers the perturbative solution obtained in section 3 (\ref{fos}-\ref{sara}).
(5) In the  ordinary geometry,
the parameter $\rho^2$ is interpreted as the size of the YM instantons. It is also
a real Kahler parameter 
of the  resolution of ADE singularities by blowing up two-spheres.
 In the noncommutative YM theory in NHS($\eta$,0),  $\rho^2$ is shifted by $\eta$ and becomes 
a quaternionic parameter $\varrho$ as shown here below
\bea
 \varrho&=&\rho^2-i\eta\nn\\       
&=&\rho^2-i u^+_{(k}u^-_{l)}\eta^{(kl)}.
\eea
 (6) Finally observe that it is possible to work out an other 
 solution of the constraint (\ref{mostapha}) by taking
\be
A=-i{1\o{\rho^2+i\eta}}.
\ee
 This choice affects (\ref{sofab}) which becomes 
\bea
&\(\D+i{\pe\o{\rho^2+i\eta}}\)\e-2\eta AB&=0\nn\\
&\(\D+i{\pe\o{\rho^2+i\eta}}\)\k&=0.
\eea
 The second class of solutions of (\ref{mostapha}) are given by (\ref{hep}) and (\ref{phe}) up to performing $\eta \to -\eta$ and $\k \leftrightarrow\e$
 . \\
\section{Conclusion}
\hspace{1cm}In this paper we have studied Yang-Mills Instantons
on noncommutative harmonic space NHS($\eta$,0). This approach has
the advantage of allowing to  explicit exact solutions of the
noncommutative self dual Yang Mills constraint eqs. It also
has the merit of going beyond the perturbative solution described in
\cite{a}.

We have first developed harmonic space noncommutative geometry
and have shown that NHS($\eta,\theta$) has two remarkable
NHS subspaces in addition to the usual ordinary one. This property
may be seen by remarking that  NHS($\eta,\theta$) has two deformation
parameters $\eta$ and $\theta$ transforming as (1,0) and (0,1)
representations of SU(2)$\times$SP(1)$\sim$ SO(4) Lorentz group.
According to whether $\eta$ and $\theta$ are  zero or not, we obtain
four subspaces, three of them are noncommutative.  Second we have
reformulated the noncommutative self-duality Yang-Mills
constraints in NHS($\eta,\theta$) by extending the idea of
harmonic analyticity. In this formulation, the basic objects
carrying the Yang-Mills self-dual constraints  are given by
the harmonic connections $\hV$
and $\hv$ of a gauged SU(2) symmetry. The latter is a generalisation of
a well known trick allowing to go from the standard analysis to the
harmonic one. The noncommutative gauge fields $\widehat {A}_M$  and
curvatures $\widehat{F}_{\mu\nu}$ are related to $\hV$ and $\hv$
as shown in eqs (\ref{mostapha}). Third we have studied the solutions
 of 
(\ref{mostapha}) by considering in a first
step
perturbative solutions around the ordinary one. In a second step,
we have given an  explicit derivation of an  exact solution of
the self-dual Yang-Mills constraints. It should be noted here
that besides the power of harmonic space analysis, this exact
solution has been made possible due to the choice of
${\theta}^{(\a\b)}=0$ and  useful properties described
in section 4.\\
\vspace{0.5cm}
\begin{center}
{\bf Acknowledgements}\\
\end{center}
Belhaj would like to thank the organizers of the Spring Workshop on Superstrings and related Matters ( March 2000), The  Abdus Salam International Centre for theoretical Physics Trieste, Italy  for hospitality ; where a part of this work is done.  He is grateful to  N. Nekrasov and S. Kachru  for valuable discussions.\\
Sahraoui would like to thank DFG under contract 445 Mar 113/5/0 to have supported his stay at the university of Muenchen, Pro. J. Wess to have invited him and Dr. R. Dick for his kindness and hospitality. Many thanks for Pro S. Theisen for his helpful discussions and suggestions.\\
\\
This research is supported by the program SARS  99/2000. contract CNCPRST-Universit\'e Mohammed V, Rabat.\\


\begin{thebibliography}{88}
\bibitem[1]{a}N. Seiberg and E. Witten, ``String Theory and Noncommutative geometry''; {\it JHEP} {\bf 9909} (1999) 032,  {\tt hep-th/9908142}. 
\bibitem[2]{b}A.  Connes, M. R. Douglas and A. Schwarz, ``Noncommutative  Geometry and Matrix theory : Compactification on torus'';
               {\it Adv. Theor. Math Phys.} {\bf 1} (1998) 35. {{\tt hep-th/9711162}}. 
\bibitem[3]{c}P.-M.  Ho, Y.-Y. Wu and Y.-S. Wu, ``Towards a Noncommutative Geometric Approach to Matrix Compactifications''; {\it Phys .Rev}.{\bf D 58}
              (1998) 026006, {\tt hep-th/9712201}.
\bibitem[4]{d} I.   Benkadour , M.  Bennai, E. Y.  Diaf and E. H.  Saidi, ``On  Matrix  model Compactification on non commutative $F_0$ geometry ''; {\it  Class . Quant . Grav} {\bf 17}(2000)
\bibitem[5]{e} N.  Nekrasov and A.  Schwarz, ``Instanton on  Non Commutative $R^4$ and (2,0) Superconformal Six dimensional theory'';  {\it Commun Math. Phys} {\bf 198}(1998) 689-703 {\tt hep-th/9802068}.
\bibitem[6]{f} H. W.  Braden, N.  Nekrasov, ``Space -Time Foam Non-Commutative Instanton'';  {\tt hep-th/9912019}.
\bibitem[7]{g} N. Nekrasov, ``lecture Notes at Spring Workshop ICTP Trieste'';  March 2000
\bibitem[8]{h} N. Seiberg and E.  Witten,``D1-D5 and singular CFT''; {\it JHEP} {\bf 9904} (1999) 017,{\tt hep-th/9903224}.
\bibitem[9]{i} O.  Aharony, M. Berkooz``IR Dynamics of d=2 , N=(4,4) Gauge Theories and DLCQ of Little String theories''; {\it JHEP}  {\bf 9910}  (1999) 0030, {\tt hep-th/9909101}.
\bibitem[10]{j} A. Belhaj . El Fallah and E.H.Saidi; {\it Class.  Quant. Grav} {\bf 16} (1999)3297 A.\\
 Belhaj . El Fallah and E.H.Saidi; {\it Class.  Quant. Grav} {\bf 17} (2000)515\\
 A.  Belhaj and E.H. Saidi, ``HyperKahler singularities in superstrings compactification and  2D N=4 Conformal Field Theory''; {\tt hep-th/0002205}
\bibitem[11]{k} C. G.  Callan, C.  Lovelace, C.R.  Nappi, S.A. Yost,``String loop corrections to Beta Functions''; {\it Nucl. Phys.}{\bf B 288}(1987) 525.
\bibitem[12]{l} E. M. Sahraoui; Proceedings of the Workshop on  ``Noncommutative Geometry, Superstring theories and Particle Physics''; 16-17 June 2000 Rabat Morocco. {\it to appear}
\bibitem[13]{m} J.  Wess, ``Lectures at Workshop  On Noncommutative Geometry,  Superstring theories  and Particle Physics''; 16-17 June 2000 Rabat Morocco.
\bibitem[14]{n} J. Madore, S. Schraml. P.Schupp and J. Wess, ``gauge theory on noncommutative spaces'';  {\tt hep-th/0001203}
\bibitem[15]{o} H. Weyl, ``Quantum mechanik und gruppen theorie'';  {\it Z. Physik} {\bf 46} (1927)1
\bibitem[16]{p} M.  Chaichain, A. Demichev and P.  Presnajder, ``Quantum field theory on Noncommutative spaces times and the persistence of ultraviolet Divergences'';  {\tt hep-th/981218.}
\bibitem[17]{q} M- Sheikh - Jabbari, ``One Loop renormalisability of Supersymmetric Yang -Mills Theories on Noncommutative torus '';  {\it JHEP} {\bf 06}(1999) 015.{\tt  hep-th/9903107} .
\bibitem[18]{r}J.  Maldacena and J.  Russo, ``Large N limit of Non-commutative Gauge theories'';  {\tt hep-th/9908134}.
\bibitem[19]{s} S.  Minwalla, M.V.  Raamsdonk and N.  Seibeg, ``Noncommutative perturbative Dynamics'';  {\tt hep-th/9912072}
\bibitem[20]{t} S. Cho, R. Hinterding, J. Madore and H. Steinacker, ``Finite field theory on noncommutative geometry'';  {\tt hep-th/9903239}.
\bibitem[21]{u} A.  Galperin, E.  Ivanov, S.  Kalitzin, V.  Ogievetsky and E.  Sokatchev; {\it Class . Quant . Grav} {\bf 1}(1984) 469.
\bibitem[22]{v} A. Galperin, Anh ky Nguyen and E. Sokatchev; {\it Class. Quant. Grav} {\bf 4}(1987)1235
\bibitem[23]{w} A. Galperin, E. Ivanov, V. Ogievetsky and E. Sokatchev; {\it Class. Quant. Grav.} {\bf 4}(1987)1255
\bibitem[24]{x} A. Galperin, E. Ivanov, V. Ogievetsky and E. Sokatchev; {\it Com. Math} {\bf 103}(1986)515
\bibitem[25]{y} E.  Sahraoui and E. H.  Saidi,`` On the completely integrable four dimensions N=2 hypermultiplet self - couplings'';  {\it Class . Quant . Grav}. {\bf 16}(1999) 1605. 
\bibitem[26]{z} A. Galperin, E. Ivanov, V. Ogievetsky and P. K. Townsend; {\it Class. Quant. Grav} {\bf 3} (1986)625
\bibitem[27]{ab}A. Galperin, E. Ivanov, V. Ogievetsky and E. Sokatchev;  {\it Annals of physics} {\bf 185}(1988)1
\bibitem[28]{ac}S. Kalitzin and E. Sokatchev; {\it Class. Quant. Grav.} {\bf 4}(1987)173
\bibitem[29]{ad}A. Galperin, E. Ivanov, V. Ogievetsky and E. Sokatchev; {\it Class. Quant. Grav.}  {\bf 2}(1985)601
\bibitem[30]{ae}J. A. Bagger, A. Galperin, E. Ivanov and V. Ogievetsky; {\it Nucl. Phys} {\bf B 303} (1988)522
\bibitem[31]{af} T.  Lhalabi and E.H. Saidi; {\it Nucl. Phys.} {\bf B 335} (1990) 689
\bibitem[32]{ag} A. Galperin, E. Ivanov, V. Ogievetsky and E. Sokatchev;  {\it Anals of Physics. } {\bf 185}(1988)22
\bibitem[33]{ah}G. W. Gibbons, D. Olivier, P. J. Ruback and G. Valent; {\it Nucl. Physi.}  {\bf B 296}(1988)679
\bibitem[34]{ai} C. N. Yang;  {\it Phys  Rev. Lett.} {\bf 38}(1977)1377
\bibitem[35]{aj}  B. Zumino; {\it Phys. lett.} {\bf B 87} (1979)203
\bibitem[36]{ak} L. Alvarez Gaum\'e and D. Z Freedman;  {\it Commun. Math. Phys.} {\bf 80}(1981)443\\
             J. Bagger and E. Witten;  {\it Nucl. Phys.} {\bf B 222} (1983)1.
\end{thebibliography}
\end{document}